\documentclass[preprint,12pt]{aastex}
\usepackage{graphicx}
\usepackage{natbib}
\newcommand{\epsB}{\epsilon_{B,-1}}
\newcommand{\epse}{\epsilon_{e,-1}}
\newcommand{\G}{\Gamma}
\newcommand{\Go}{\Gamma_0}
\newcommand{\bo}{\beta_0}
\newcommand{\be}{\beta}
\newcommand{\g}{\gamma}
\newcommand{\nuo}{\nu_{obs}}
\newcommand{\td}{t_{dec}}

\title{Radio Remnants of Compact Binary Mergers - the Electromagnetic Signal that will follow the Gravitational Waves}

\author{Ehud Nakar$^{1}$ and Tsvi Piran$^{2}$}
\affil{$^1$\ Raymond and Beverly Sackler School of Physics \&
Astronomy, Tel Aviv University, Tel Aviv 69978, Israel;
udini@wise.tau.ac.il} \affil{$^2$\ Racah Institute of Physics, The
Hebrew University, Jerusalem 91904, Israel; tsvi@phys.huji.ac.il}

\begin{document}


\begin{abstract}
The question ``what is the observable electromagnetic (EM) signature
of a compact binary  merger?" is an intriguing one with  crucial
consequences to the quest for  gravitational waves (GW). Compact
binary mergers  are prime sources of GW, targeted by current and
next generation detectors. Numerical simulations have demonstrated
that these mergers eject energetic sub-relativistic (or even
relativistic) outflows. This is certainly the case if the mergers
produce short GRBs, but even if not, significant outflows are
expected. The interaction of such outflows with the surround matter
inevitably leads to a long lasting radio signal. We calculate the
expected signal from these  outflows (our calculations are also
applicable to short GRB orphan afterglows)  and we discuss their
detectability. We show that the optimal search for such signal
should, conveniently, take place around 1.4 GHz. Realistic estimates
of the outflow parameters yield signals of a few hundred $\mu$Jy,
lasting a few weeks, from sources at the detection horizon of
advanced GW detectors. Followup radio observations, triggered by GW
detection, could reveal the radio remnant even under unfavorable
conditions. Upcoming  all sky surveys can detect a few dozen, and
possibly even thousands, merger remnants at any give time, thereby
providing robust merger rate estimates even before the advanced GW
detectors become operational. In fact, the radio transient RT
19870422 fits well the overall properties predicted by our model and
we suggest that its most probable origin is a compact binary merger
radio remnant.

\end{abstract}

\section{Introduction}
\label{sec:introduction} Compact binary  (Neutron star - Neutron
star, NS$^2$, or Black Hole - Neutron star, BH-NS) mergers are  prime
sources of gravitational radiation. The GW detectors LIGO
\citep{LIGO09},  Virgo \citep{Virgo08} and GEO600 \citep{GEO2008}
are designed to optimally detect merger signals. These detectors
have been operational intermittently during the last few years
reaching their nominal design sensitivity
\citep{LIGO09a,LIGOVirgo2010,LIGOVirgo2010a} with  detection
horizons of   a few dozen Mpc for NS$^2$ and almost a hundred Mpc for
BH-NS mergers (the LIGO - Virgo collaboration adopts an optimal
canonical distance of 33/70Mpc;  \citealt{LIGORate2010}). Both LIGO
and Virgo are being upgraded now and by the end of 2015 are expected
to be operational at  sensitivities $\sim 10-15$ times greater than
the initial LIGO \citep{AdvancedLIGO2009}, reaching  a few hundred
Mpc detection horizon  for NS$^2$ mergers  and a Gpc for BH-NS
mergers (445/927 Mpc are adopted by the LIGO-Virgo collaboration as
canonical values; \citealt{LIGORate2010}).

Understanding the observable EM signature of compact binary mergers
has several observational implications. First, once the detectors
are operational it is likely that the first detection of a GW signal
will be around or even below threshold. Detection of an accompanying
EM signal will confirm the discovery, thereby increasing
significantly the sensitivity of GW detectors \citep{KP93}. Second,
the physics that can be learned from observations of a merger event
through different glasses is much greater than what we can learn
through EM or GW observations alone. Finally, even before the
detectors are operational, detection of EM signature will enable us
to determine the expected rates, a question of outmost importance
for the design and the operation policy of the advanced detectors.

The current constraints on the rates are rather loose. The last LIGO
and Virgo runs provided only weak upper limits on the merger rates:
8700 Myr$^{-1}$($10^{10}L_\odot)^{-1}$ corresponding to $\sim 10^5$
yr$^{-1}$ Gpc$^{-3}$ for NS$^2$ and $4.5 \times 10^4$ Myr$^{-1}$
($10^{10}L_\odot)^{-1}$  ($\sim 5 \times 10^5 {\rm
~Gpc^{-3}~yr^{-1}}$) for BH-NS \citep{LIGO09a}. Estimates based on
the observed binary pulsars in the Galaxy are highly uncertain, with
values ranging from $ 20 - 2 \times 10^4  {\rm ~Gpc^{-3}~yr^{-1}}$
\citep{Phinney91,NPS91,KalEtal04,KalEtal04a,LIGORate2010}. It has
been suggested \citep{Eichler89} that short Gamma-Ray Bursts (GRBs)
arise from neutron star merger events. The estimated  rate of short
GRBs are indeed comparable to binary pulsar estimates
\citep{GP06,NGF06,GS09}. However, while appealing, the association
is not proven yet \citep{Nakar07}. If correct, the observed rate of
short GRBs, $\sim 10  {\rm ~Gpc^{-3}~yr^{-1}}$,  provides a lower
limit to the merger rate. The true rate depends on a poorly
constraint beaming angle, resulting in an uncertainty of almost two
orders of magnitude. There are no direct estimates of BH-NS mergers,
as no such system has ever been observed, and here one has to relay
only on a rather model dependent population synthesis
\citep[e.g.,][]{BKR+08,MO10}.


Possible EM signals  from coalescence events were discussed by
several authors. There are several suggestions
\citep{HL01,mk04,pp10} of a prompt (coinciding with the GW signal)
short lived EM signals, mostly in low radio frequencies,
whose amplitudes are highly uncertain. \cite{LP98} suggested that
the radioactive decay of ejected debris from the merger will drive a
short lived supernova like event. \cite{Metzger+10} calculated the
radioactive heating during this process self-consistently. They find
that if $0.01 {\rm M_\odot}$ is ejected then the optical emission
from a merger at $300$ Mpc peaks after $\sim 1$ day at $m_V \approx
24$. If the mass ejection is lower then the optical emission will be
even fainter. Finding, and especially identifying the origin of,
such rare and faint events in the crowded variable optical sky is an
extremely challenging task, even for current and future optical
searches like PTF \citep{Law09,Rau09}, PanSTARR, and LSST
\citep{LSST09}.

An intriguing possibility is that mergers produce short GRBs
\citep{Eichler89}. However, short GRBs are expected to be beamed,
and  only rarely this EM signal will point towards us. A beamed GRB
that is observed off-axis will inevitably produce a long lasting
radio ``orphan" afterglow
\citep{Rhoads97,Waxman+98,Frail+00,Levinson+02,Gal-Yam+06}. A key
point in estimating the detectability of GRB orphan afterglows is
that the well constrained observables are the {\it isotropic}
equivalent energy of the flow and the rate of bursts that point
towards earth. However, the detectability of the orphan afterglows
depends only on the {\it total} energy and  {\it true} rate, namely
on the poorly constraint jet beaming angle. \cite{Levinson+02} have
shown that while large beaming increases the true rate it reduces
the total energy, and altogether reduces the detectability of radio
orphan afterglows. This counterintuitive result makes the
detectability of late emission from a decelerating jet, which
produced a GRB when it was still ultra-relativistic, less promising.
However, regardless of amount of ulrtra-relativistic outflow that is
launched by compact binary mergers, and of whether they produce
short GRBs or not, mergers are most likely do launch an energetic
sub-relativistic and mildly-relativistic outflows. The interaction
of these outflows with the surrounding matter will inevitably
produce blast waves and possibly stronger radio counterparts than
that of ultra-relativistic outflows.

Below we first discuss (in \S 2) the current estimates of mass and
energy ejection from compact binary mergers. In \S 3 we calculate
the radio emission resulting from the interaction of this ejecta
(sub-relativistic, mildly relativistic and off-axis relativistic)
with the surrounding interstellar matter (ISM). The calculations
follow to a large extent models of radio Ib/c
supernovae\footnote{Note that while blast waves from radio SNe
propagate into a surrounding wind density profiles the blast wave
from a merger is expected to propagate into a constant density
surrounding matter.} \citep{Chevalier98,CF06} and long GRB radio
afterglows \citep{SPN98,Waxman+98,GranotSari02}. The success of
radio supernova (SN) modeling, where the observations are superb,
indicates that the microphysics is well constrained ant that
equipartition parameters describe well the physical conditions.
Hence the main uncertainty in the predicted radio signal is in the
amount of matter ejected from the merger event and its velocity.
Luckily the estimates of this important quantity can be
significantly improved even using existing numerical models.

We discuss the observational implications for detectability of
merger remnants in \S 4. We estimate the expected rates of detection
of different outflows in \S 4.1. We devote in \S 4.2 a special
attention to short GRB  orphan afterglows that are a special case of
our model, in which the outflow is launched relativistically, but
the radio emission peaks only during its mildly relativistic phase.
The estimates of orphan afterglows detectability  is independent of
whether they are the products of binary mergers or not. In \S 4.3 we
examine possible other radio transients that may hinder the
identification of merger remnants. Finally in \S 4.4 we examine
blind transient searched done in the past and we identify RT
19870422 as a possible and even likely merger remnant. We conclude
in \S 5.

\section{Mass and energy ejection from compact binary mergers}

Numerical simulations of compact binary mergers have been carried
out by various groups with two different approaches. Some
\citep[e.g.,][]{RDTP00,RJ01,RP07}  use Newtonian   dynamics
(modified to allow for gravitational radiation back-reaction) with
detailed microphysics.
Others  \citep[e.g.,][]{yst08,RBGLF10} use full  General
relativistic dynamics with different levels of approximate
microphysics,
with or without MHD. In almost all NS$^2$ simulations one finds an
accretion disk surrounding a rapidly rotating massive object that
eventually collapses to a black hole. An exception is the recent
general relativistic simulations of \cite{KSST10} who find no disks
in some configurations. The system lifetime is  at least a few dozen
milliseconds and possibly longer. The fate of an accretion disk in a
BH-NS merger is expected to depend on the mass ratio, the BH spin
and the NS compactness. In some cases the disk is very small while
in others it is substantial \citep[e.g.,][]{Faber06,Shibata08}.

All simulations find some form of relativistic or sub-relativistic
mass ejection. First, matter is ejected as tidal tails during the
first stages of the merger. In BH-NS mergers the ejected energy can
be very high and its velocity is mildly relativistic. For example
\cite{Rosswog05} find $0.5$ c ejecta with $\sim 10^{52}$ erg, where
c is the light speed. In a NS$^2$ mergers a lower, but yet
significant, amount of energy can be ejected (e.g.,
\citealt{Rosswog+99} find $\sim 10^{51}$ erg) at a lower velocities
of $0.1-0.2$ c. This mass ejection is expected also if no
significant disk is formed. Disk formation leads to several
additional outflow sources. First, neutrino heating drives a wind
from the disk surface \citep[e.g.,][]{levinson06,Metzger+08,d+09}.
The energy in this wind is substantial with predictions ranging
between $10^{49}$ erg and $10^{51}$ erg for 0.01-0.1 M$_\odot$ disk.
The outflow velocity is $\sim 0.1-0.2$ c from the outskirts of the
disk and it is increasing, possibly up to relativistic velocities,
for wind that is ejected from close to the central object. The mass
ejection becomes even stronger when neutrino heating shuts-off and
the wind is driven by viscous heating and by He-Synthesis
\citep{Lee+09,Metzger+09}, leading to an ejection of $20-50\%$ of
the initial disk mass at $0.1-0.2$ c. Additional energy source is
neutrino-antinuetrino annihilation above the disk, which can deposit
up to $\sim 10^{49}$ erg, into an amount of mass that is not well
constrained, leading possibly to a relativistic outflow. Finally,
more speculative, but yet very plausible, source of outflow are EM
processes that tap the rotational energy of the central object, such
as the Blandford-Znajek Mechanism \citep{BZ77}. These are likely to
produce relativistic outflows with an energy that can be as high as
$10^{52}$ erg, and are the most probable engines of short GRBs, if
those are produced by compact binary mergers.

The conclusion is that a significant mass and energy ejection is a
prediction of almost all compact binary merger modelings. In NS$^2$
mergers an ejection of $\gtrsim 10^{50}$ erg at $0.1-0.2$ c is a
fairly robust prediction. Faster ejecta (relativistic or mildly
relativistic) with energy $\gtrsim 10^{49}$ is also quite likely
from inner parts of the 0.01-0.1 M$_\odot$ disk that is typically
found in simulations. The outflow from BH-NS mergers was explored
only by a few authors, but it is also seems to be significant and
potentially even more energetic and at faster velocities than the
outflow from NS$^2$ mergers.

\section{The radio signal from outflow-ISM interaction}\label{sec Theory}

Consider a spherical outflow with an energy $E$ and an initial
Lorentz factor $\Go$, with a corresponding velocity $c \beta_0$,
that propagates into a constant density, $n$, medium. If the outflow
is not ultra relativistic, i.e., $\Go-1 \lesssim 1$  it propagates
at a constant velocity until, at $t_{dec}$, it reaches radius
$R_{dec}$, where it collects a comparable mass to its own:
\begin{equation}
    R_{dec} = \left( \frac{3E}{4\pi n m_p c^2 \bo^2} \right)^{1/3} \approx 10^{17}
    {\rm ~cm~} E_{49}^{1/3} n^{-1/3} \bo^{-2/3},
\end{equation}
and
\begin{equation}
    t_{dec}= \frac{R_{dec}}{c\beta_0} \approx 30 {\rm ~day~}
    E_{49}^{1/3} n^{-1/3} \bo^{-5/3} ,
    \label{eq:tdec}
\end{equation}
where we approximate $\Go-1 \approx \bo^2$ and ignore relativistic
effects. Here and in the following, unless stated otherwise, $q_x$
denotes the value of $q/10^x$ in c.g.s. units. At a radius
$R>R_{dec}$ the flow decelerates assuming the Sedov-Taylor
self-similar solution, so the outflow velocity can be approximated
as:
\begin{equation}\label{eq Gamma}
    \be \approx \bo \left\{\begin{array}{cc}
                   1 & R \leq R_{dec}~,  \\
                   \left({R}/{R_{dec}}\right)^{-3/2} & R \geq R_{dec} ~.
                \end{array} \right.
\end{equation}

If the outflow is collimated, highly relativistic and points away
from a generic observer, as will typically happen if  the mergers
produce short GRBs, the emission during the relativistic phase will
be suppressed by relativistic beaming. Observable emission is
produced only once the external shock decelerates to mildly
relativistic velocities and the blast-wave becomes quasi spherical.
This takes place when $\Gamma \approx 2$ namely at $R_{dec}(\bo =
1)$. From this radius the hydrodynamics and the radiation become
comparable to that of a spherical outflow with an initial Lorentz
factor $\G_0 \approx 2$. This behavior is the source of the late
radio GRB orphan afterglows \citep{Rhoads97,Levinson+02}. Our
calculations are therefore applicable for the detectability of
mildly and non-relativistic outflows as well as  for radio orphan
GRB afterglows.

Emission from Newtonian and mildly relativistic shocks is observed
in radio SNe and late phases of GRB afterglows. These observations
are well explained by a theoretical model involving synchrotron
emission of shock accelerated electrons in an amplified magnetic
field.  The success of this model in explaining the detailed
observations of radio Ib/c SNe
\citep[e.g.,][]{Chevalier98,Soderberg+05,CF06} allows us to employ
the same microphysics here. Energy considerations show that both the
electrons and the magnetic field carry significant fractions of the
total internal energy of the shocked gas, $\epsilon_e \approx
\epsilon_B \sim 0.1$ . These values are consistent with those
inferred from late radio afterglows of long GRBs
\cite[e.g.,][]{Frail+00,Frail+05}. The observed spectra indicate
that the distribution of the accelerated electrons Lorentz factor,
$\g$,  is a power-law $dN/d\g \propto \g^{-p}$ at $\g>\g_m$ where $p
\approx 2.1-2.5$ in mildly relativistic shocks (e.g., the radio
emission from GRB associated SNe and late GRB afterglows) and $p
\approx 2.5-3$ in Newtonian shocks (as seen in typical radio SNe;
\citealt[][and references therein]{Chevalier98}). The value of
$\g_m$ is not observed directly but it can be calculated based on
the total energy of the accelerated electrons, $\g_m =
\frac{p-2}{p-1} \frac{m_p}{m_e} \epsilon_e \be^2$

The radio spectrum generated by the shock is determined by two
characteristic frequencies\footnote{The cooling frequency is
irrelevant in the radio.}. One is
\begin{equation}\label{eq num}
    \nu_m \approx 1 {\rm ~GHz~} n^{1/2} \epsB^{1/2} \epse^{2} \be^{5} ,
\label{eq:num}
\end{equation}
the typical synchrotron frequency of electrons with the typical
(also minimal) Lorentz factor $\g_m$. The other is $\nu_a$, the
synchrotron self-absorption frequency. We show below that since we
are interested at the maximal flux at a given observed frequency,
$\nu_a$ may play a role only if it is larger than $\nu_m$. Its value
in that case is
\begin{equation}\label{eq nua}
    \nu_a (>\nu_m) \approx 1 {\rm ~GHz~} R_{17}^\frac{2}{p+4}n^\frac{6+p}{2(p+4)} \epsB^\frac{2+p}{2(p+4)} \epse^\frac{2(p-1)}{p+4}
    \be^\frac{5p-2}{(p+4)} .
\label{eq:nua}
\end{equation}

\begin{figure}[t]
  \includegraphics[width=13cm]{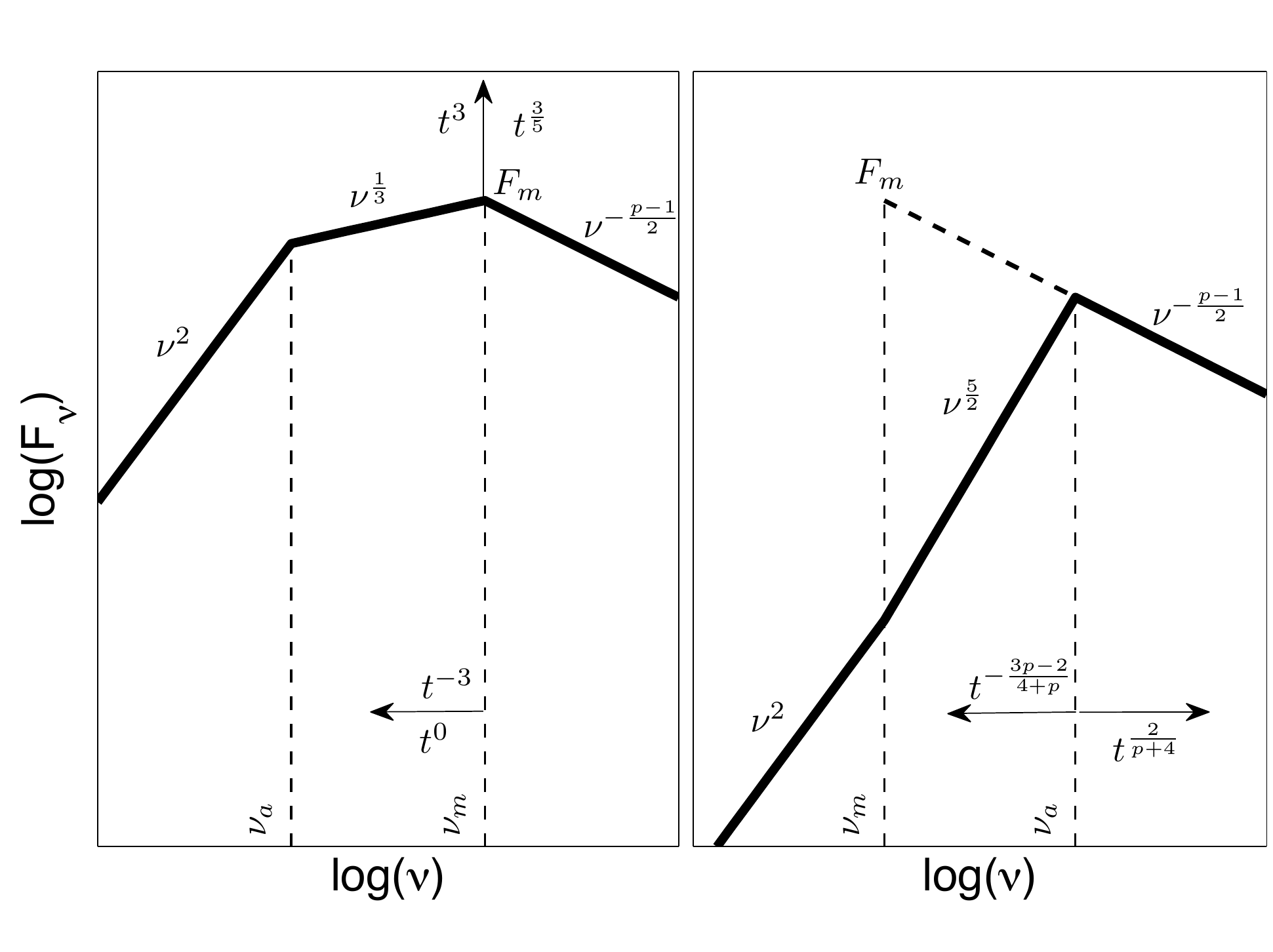}\\
  \caption{A schematic sketch of the two possible spectra and the evolution of the characteristic flux, $F_m$, and frequencies, $\nu_a$ and $\nu_m$.
  The arrows show the temporal evolution of the characteristic values. The temporal dependence before $t_{dec}$ is
  noted below/to the left of the arrows while the temporal dependence after $t_{dec}$ is
  noted above/to the right of the arrows. Note that the evolution of $\nu_m$ and $F_m$, marked only in the left spectrum,
  is relevant for both spectra. The evolution of $\nu_a$, marked only in the right spectrum, is correct only when $\nu_m<\nu_a$ and is
  therefore relevant only in that spectrum.}\label{fig: spec}
\end{figure}

Figure \ref{fig: spec} illustrates the two possible spectra,
depending on the order of $\nu_a$ and $\nu_m$. The flux at any
frequency can be found using these spectra and  the unabsorbed
synchrotron flux at $\nu_m$:
\begin{equation}\label{eq Fm}
    F_m \approx 0.5 {\rm ~mJy~}  R_{17}^3 n^{3/2} \epsB^{1/2}
    \be d_{27}^{-2} ,
\end{equation}
where $d$ is the distance to the source (we neglect any cosmological
effects). Note that this is the real flux at $\nu_m$ only if $\nu_a
< \nu_m$ (see figure \ref{fig: spec}).

As long as the shock is moving with a constant velocity  i.e., at $t<t_{dec}$, the
flux across the whole spectrum increases (see
equations \ref{eq num}-\ref{eq Fm}).
The flux evolution at later times depends on the spectrum at
$t_{dec}$, namely on
\begin{equation}\label{eq numdec}
    \nu_{m,dec} \equiv \nu_m(t_{dec}) \approx 1 {\rm ~GHz~} n^{1/2} \epsB^{1/2} \epse^{2}
    \bo^{5} ,
\end{equation}
 and if $\nu_{a,dec}>\nu_{m,dec}$ then possibly on
\begin{equation}\label{eq nuadec}
 \nu_{a,dec} \equiv  \nu_a(t_{dec})  \approx 1 {\rm ~GHz~} E_{49}^\frac{2}{3(4+p)} n^\frac{14+3p}{6(4+p)} \epsB^\frac{2+p}{2(4+p)} \epse^\frac{2(p-1)}{4+p}
    \bo^\frac{15p-10}{3(4+p)} .
\end{equation}
The flux at that time can be found using the unabsorbed synchrotron flux at $\nu_{m,dec}$:
\begin{equation}
    F_{m,dec} \approx 0.5 {\rm ~mJy~} E_{49} n^{1/2} \epsB^{1/2}
    \bo^{-1} d_{27}^{-2} .
\end{equation}

\begin{figure}[h]
  \includegraphics[width=12cm]{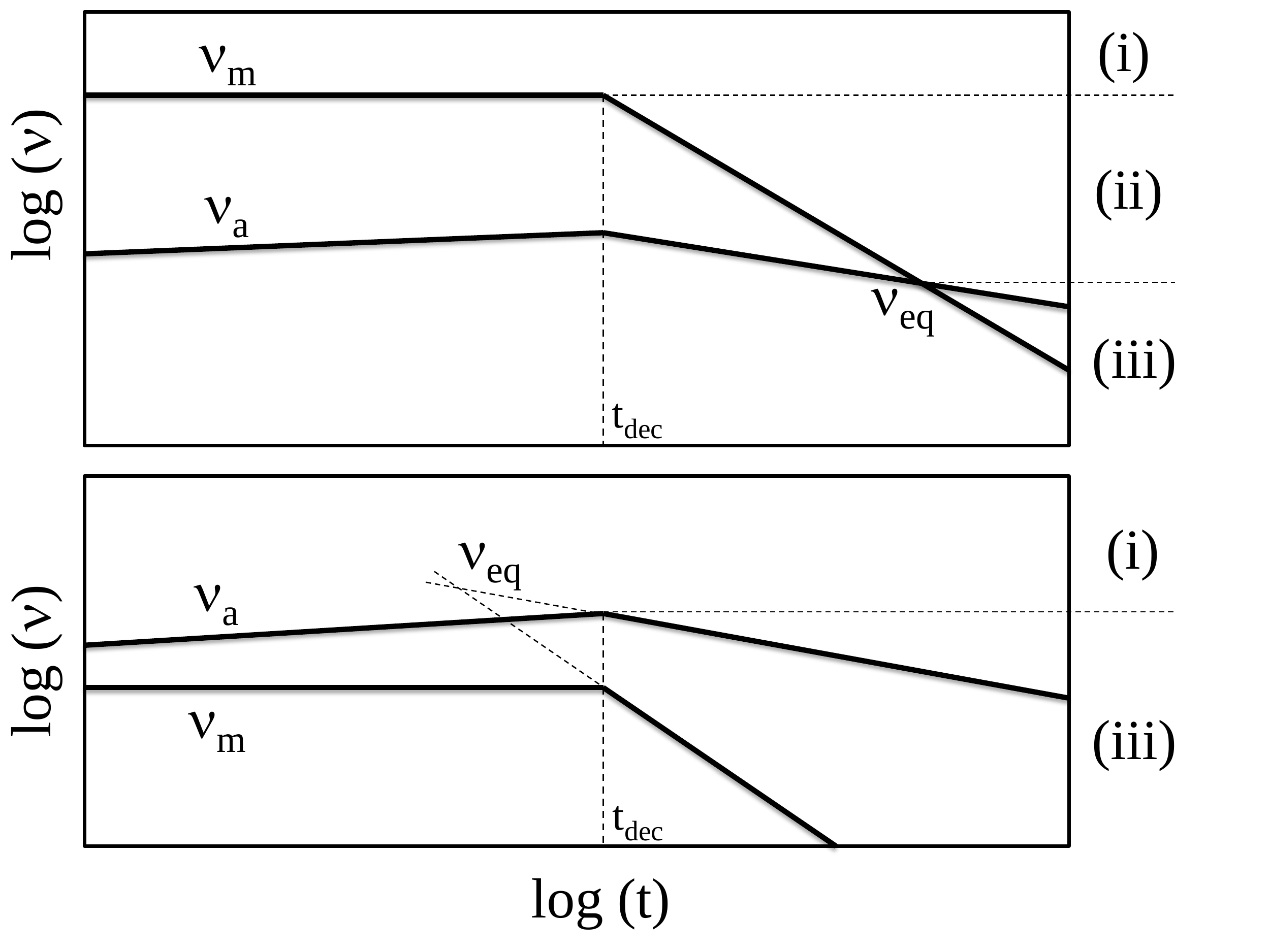}\\
  \caption{A
schematic sketch of the time evolution of $\nu_a$ and $\nu_m$ in two
cases, $\nu_{a,dec}<\nu_{m,dec}$ (top) and $\nu_{a,dec}>\nu_{m,dec}$
(bottom). Also marked is the value of $\nu_{eq}$. The vertical
dashed line marks $\td$. The ranges of $\nuo$ at which each of the
cases is observed is separated by horizontal dashed lines and marked
on the right. Note that in the bottom panel $\nu_m$ and $\nu_a$ are
not crossing each other at $t>\td$ and only two types of light
curves, cases (i) and (iii), can be observed.}\label{fig:cases}
\end{figure}

Consider now a given observed frequency $\nuo$. We are interested at
the light curve near the peak flux at this frequency. There are
three possible types of light curves near the peak corresponding to:
(i) $\nu_{m,dec},\nu_{a,dec}<\nuo$, (ii) $\nu_{eq}< \nuo <
\nu_{m,dec}< $ and (iii) $ \nuo < \nu_{eq}, \nu_{a,dec} $.  Where we
define
\begin{equation}
    \nu_{eq} = 1 {\rm ~GHz~} E_{49}^{1/7} n^{4/7} \epsB^{2/7}
    \epse^{-1/7}
\end{equation}
as the  frequency at which\footnote{Note that if $\nu_{m,dec} <
\nu_{a,dec}$ this equality will never take place. In that case
$\nu_{eq}$ is the frequency at which this equality would have
happened if $\Go$ would have been large enough (see figure
\ref{fig:cases})} $\nu_m=\nu_a$. In figure \ref{fig:cases} we show a
schematic sketch of the time evolution of $\nu_a$ and $\nu_m$ and
the corresponding ranges of $\nuo$ in which each of the cases is
observed.

To estimate the time and value of the peak flux we recall, that at
all frequencies  the flux increases until $t_{dec}$. In case (i),
$\nu_{m,dec},\nu_{a,dec}<\nuo$, the deceleration time, $\td$, is
also the time of the peak. The reason is that while $F_m$ increases,
$\nu_m$ decreases fast enough so that $F_{\nuo}$ decreases after
$t_{dec}$. Note that in that case $\nu_a$ plays no role since it
decreases after deceleration. Overall, in this case the flux peaks
at $t_{dec}$ and $F_{{\nuo},peak}= F_{m,dec}
(\nuo/\nu_{m,dec})^{-(p-1)/2}$.

\begin{figure}[!h]
  \includegraphics[width=12cm]{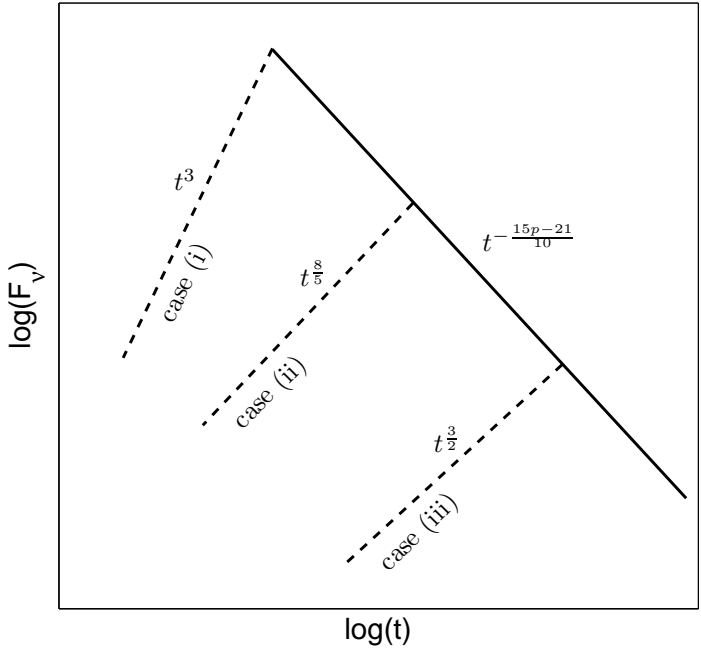}
  \caption{Schematic light curves of the three cases. The rising phase,
  marked in dashed line for each of the the phases, is that of the last
  temporal power law segment before the peak. After the peak all cases
  show the same power-law decay. }\label{fig:lightcurves}
\end{figure}

In the two other cases, (ii) and (iii), $ \nuo < \nu_{m,dec}$ and/or
$\nuo<\nu_{a,dec}$  and the flux keeps rising at $t>\td$ until
$\nuo=\nu_{m}(t)$ or $\nuo=\nu_{a}(t)$, whichever comes last.
To find out which one of the two frequencies is it, we compare
$\nuo$ with $\nu_{eq}$. At $t>t_{dec}$,  $\nu_m$ decreases faster
than $\nu_a$. Therefore in case (ii) where $\nu_{eq}<\nuo$,  the
last frequency to cross $\nuo$ is $\nu_m$ and the peak flux is
observed when $\nuo=\nu_m(t)$. In case (iii) where $\nuo<\nu_{eq}$,
and the last frequency to cross $\nuo$ is $\nu_a$ and the peak flux
is observed when $\nuo=\nu_a(t)$. Now, it is straight forward to
calculate the peak flux, $F_{{\nuo},peak}$ and the time that it is
observed, $t_{peak}$, for different frequencies. It is also straight
forward to calculate the flux temporal evolution prior and after
$t_{dec}$ using equations \ref{eq num}-\ref{eq Fm} and the relation
$t \propto R$ which holds at $t<t_{dec}$ and $\be \propto t^{-3/5}$
at $t>t_{dec}$. The peak fluxes, the times of the peak and the
temporal evolution of the three different cases are summarized in
table \ref{table1}. The overall different light curves are depicted
in Fig. \ref{fig:lightcurves}

\begin{table}[h]
\begin{tabular}{|l|c|c|c|c|}
   \hline
  Regime & $F_{{\nuo},peak}/F_{m,dec}$ & $t_{peak}/t_{dec}$ & $F_{\nuo}^\dag$ & $F_{\nuo}$ \\
  &  &  & $t<t_{peak}$ & $t_{peak}<t$ \\
  \hline\hline
  $\nu_{m,dec},\nu_{a,dec}<\nuo$ & $
\left({\nuo}/{\nu_{m,dec}}\right)^{-\frac{p-1}{2}}$ & $1$ & $\propto
t^{3}$ & $\propto t^{-\frac{15p-21}{10}}$  \\
\hline
  $\nu_{eq}<\nuo<\nu_{m,dec}$ & $
\left({\nuo}/{\nu_{m,dec}}\right)^{-1/5}$ &
$({\nuo}/{\nu_{m,dec}})^{-1/3}$ & $\propto t^{\frac{8}{5}}$ &
$\propto t^{-\frac{15p-21}{10}}$  \\
\hline
  $\nuo<\nu_{eq},\nu_{a,dec}$ & $
~\nu_{m,dec}^\frac{p-1}{2}
~\nu_{a,dec}^{-\frac{3(p+4)(5p-7)}{10(3p-2)}}
~\nuo^\frac{(32p-47)}{5(3p-2)}$ &
$({\nuo}/{\nu_{a,dec}})^{-\frac{4+p}{3p-2}}$ & $\propto
t^{\frac{3}{2}}$ & $\propto t^{-\frac{15p-21}{10}}$  \\
  \hline
\end{tabular}
\caption{\label{table1} The observed flux before and after
$t_{peak}$ in the three different regimes. \newline $\dag$ The
temporal evolution only during the last power-law segment before
$t_{peak}$. At earlier times the temporal evolution may be
different.}
\end{table}

The most sensitive radio facilities are at frequencies of $1.4$ GHz
and higher. Equations \ref{eq numdec} and \ref{eq nuadec} imply that
in this frequency range, for most realistic scenarios, it is a case
(i) light curve, i.e., $\nu_{a,dec},\nu_{m,dec}<\nuo$. Therefore, a
Newtonian and mildly relativistic outflows as well as relativistic
GRB orphan afterglows peak at $t_{dec}$ with:
\begin{equation}\label{eq Fpeak reg1}
    F_{\nuo,peak}(\nu_{a,dec},\nu_{m,dec}<\nuo) \approx 0.3 {\rm ~m Jy~} E_{49} n^\frac{p+1}{4} \epsB^\frac{p+1}{4}
    \epse^{p-1} \bo^\frac{5p-7}{2} d_{27}^{-2} \left(\frac{\nuo}{1.4 {\rm
    ~GHz~}}\right)^{-\frac{p-1}{2}} .
\end{equation}

The regime of $F_{\nuo,peak}$ at lower radio frequencies ($< 1$ GHz)
depends on the various parameters. If the outflow is Newtonian or
the density is low or the energy is low then
$\nu_{a,dec},\nu_{m,dec} < 100$ MHz and equation \ref{eq Fpeak reg1}
is applicable. Otherwise low radio frequencies are in regime (iii),
i.e., $\nuo<\nu_{eq},\nu_{a,dec}$. The flux peaks in this case at
\begin{equation}\label{eq tpeak reg3}
    t_{peak}(\nuo<\nu_{eq},\nu_{a,dec}) \approx 200 {\rm ~day~} E_{49}^\frac{5}{11}
    n^\frac{7}{22} \epsB^\frac{9}{22} \epse^\frac{6}{11} \left(\frac{\nuo}{150 {\rm
    ~MHz~}}\right)^{\frac{13}{11}} ,
\end{equation}
with
\begin{equation}\label{eq Fpeak reg3}
    F_{\nuo,peak}(\nuo<\nu_{eq},\nu_{a,dec}) \approx 50 {\rm ~\mu Jy~} E_{49}^\frac{4}{5} n^\frac{1}{5} \epsB^\frac{1}{5}
    \epse^\frac{3}{5} d_{27}^{-2} \left(\frac{\nuo}{150 {\rm
    ~MHz~}}\right)^{\frac{6}{5}} .
\label{eq:fpeaklow}
\end{equation}
In the last two equations we used  $p=2.5$ (other $p$ values in the
range 2.1-3 yield slightly different numerical factors and power
laws).

To date, the best observed signal from a mildly relativistic blast
waves is the radio emission that follows GRB associated SNe. The
main difference is that in these cases the circum burst medium is
typically a wind (i.e., $n \propto R^{-2}$) and therefore the
density at early times is much larger then in the ISM and self
absorption plays the main role in determining the light curve. A
good example for comparison of equation \ref{eq Fpeak reg1} with
observations is the light curve of SN 1998bw. This light curve is
observed at several frequencies at many epochs, enabling a detailed
modeling that results in tight constraints of the blast wave and
microphysical parameters. \cite{LiChevalier99} find that at the time
of the peak at $1.4$ GHz, about 40 days after the SN, taking
$\epsilon_e=\epsilon_B=0.1$, the energy in the blast wave is $\sim
10^{49}$ erg, its Lorentz factor is $\sim 2$ and the external
density at the shock radius is $n \sim 1 {\rm ~cm^{-3}}$. The peak
is observed when $\nu_m,\nu_a \leq \nuo$ and it depends only on
these parameters (it is only weakly sensitive to the density
profile). Therefore, equation \ref{eq Fpeak reg1} is applicable in
that case. Indeed, plugging these numbers into equation \ref{eq
Fpeak reg1} we obtain a flux of 20 mJy at the distance of SN 1998bw
(40 Mpc), compared to the observed flux of 30 mJy. This is not
surprising given that the model we use is based on that of radio
SNe.

In the discussion above we considered an outflow with a
characteristic energy and velocity. However a compact binary merger
may produce an outflow with an energy dependent velocity, e.g.,
$E(\Go) \propto (\Go-1)^{-\eta}$. If we consider a case (i) signal
($\nu_{m,dec},\nu_{a,dec}<\nuo$; e.g., above $1$ GHz) then we find
$F_{\nuo,peak} \propto E (\Go-1)^{{(5p-7)}/{4}} \propto
(\Go-1)^{(5p-7-4\eta)/4}$. Therefore if  $\eta< \frac{5p-7}{4}
\approx 1.5$ the flux is dominated by the mildly relativistic
ejecta, assuming that the relativistic part of the outflow is not
pointing towards the observer. Otherwise the flux is dominated by
the slowest ejecta.

\section{Detection, identification and possible candidates}
\subsection{detectability}
For canonical parameters the strongest signal is expected around
$1.4$ GHz, conveniently where the sensitivity of radio telescopes is
high. If the signal peaks at lower frequencies  it decreases from
the peak as $\nu^{-(p-1)/2}$. Since 1.4 Ghz receivers are ten times
more sensitive than lower frequency ones they are still more likely
to detect a signal even in this case. Therefore, $1.4$ GHz is the
optimal frequency to look for radio remnants of compact binary
mergers and we therefore consider here the delectability at a $1.4$
GHz survey.

The number of events in a single  whole sky snapshot is $N_{all-sky}
={\cal R} V \Delta t $, where $V$ is the detectable volume at the
survey flux limit $F_{lim}$, $\Delta t$ is the time that the flux is
above the detection limit and ${\cal R}$ is the event rate. Since at
$1.4$ GHz the relevant light curve is case (i) (see \S \ref{sec
Theory}) we use equations \ref{eq:tdec} and \ref{eq Fpeak reg1}, and
the approximation $\Delta t \approx t_{dec}$, to find that the
number of radio coalescence remnants in a single $1.4$ GHz whole sky
snapshot is
\begin{equation}\label{eq rate}
    N_{all-sky}(1.4 {\rm~ GHz}) \approx 20 E_{49}^{11/6} n^\frac{9p+1}{24} \epsB^\frac{3(p+1)}{8}
    \epse^\frac{3(p-1)}{2} \bo^\frac{45p-83}{12} {\cal R}_{300}
    F_{lim,-1}^{-3/2} ~.
\end{equation}
Where $F_{lim,-1}=F_{lim}/0.1$ mJy and ${\cal R}_{300}$ is the
merger rate in units of $300 {\rm ~Gpc^{-3}~ yr^{-1}}$.


Since the microphysical parameters are reasonably constrained by
radio SNe, the main uncertainty in the signal detectability is the
blast-wave properties and the circum-merger density. The latter is
expected to vary significantly between different merger events,
dropping down to $n \sim 10^{-6} {\rm~ cm^{-3}}$ for mergers that
take place outside  of their host galaxies. However, the observed
Galactic double-NS population reveals that a significant fraction of
compact binary mergers (at least NS$^2$) take place in the disks of
Milky-way-like galaxies where $n \sim 1 {\rm~ cm^{-3}}$.

As discussed in \S 2 there is a good chance that NS$^2$ mergers
eject mildly relativistic outflows with $E \gtrsim 10^{49}$ erg. In
that case the EM counterparts of these GW sources are detectable by
current facilities. A deep radio survey that covers a significant
fraction of the sky will most likely detect their radio remnants to
a distance of $\sim 300$ Mpc. These will appear as transients,
varying on a several weeks time scale, with an optically thin
spectrum at $\nu \gtrsim 1$ GHz at all time. The radio remnants will
be identified in random places within their host galaxies, which
should be easily detectable at that distance. It will not be
accompanied by any optical counterpart with similar variability time
scales. Given a GW trigger with localization of $10-100
{\rm~deg^2}$, a deep search of the error box will detect a remnant
in cases that the surrounding density is not very low.

NS$^2$ mergers are also expected to eject energetic, $E \gtrsim
10^{50}$ erg, outflow at lower velocities, $\beta_0 = 0.1-0.2$. The
detectability is very sensitive to the velocity, $N \propto
\beta_0^{2.5}$ for $p=2.5$, and to the energy, $N \propto E^{11/6}$.
Therefore, even if there is no mildly relativistic component to the
outflow, the number of detectable remnants that are dominated by
slow moving ejecta is expected to be significant. For example,
assuming that an energy of $10^{50}$ erg is ejected at $\beta_0 =
0.2$ we expect $N_{all-sky} \sim 10 {\cal R}_{300}
F_{lim,-1}^{-3/2}$. The variability time scale of these transients
is expected to be $\sim 3$ yr. If the velocity is instead $0.1$c the
number of events drops by an order of magnitude and the variability
timescale increases to $\sim 10$ yr. These time scales increase the
difficulty in the identification of the remnants as transients, and
require a long term survey. The spectrum of these transients will be
optically thin also at low frequencies ($\sim 100$ MHz). Other
characteristics of these transients (e.g., location within the host)
are expected to be similar to that of a mildly relativistic outflow.
A GW triggered search increases of course the probability to find a
remnant and a $>10^{50}$ erg outflow can be detected up to $300$ Mpc
even if its velocity is $\sim 0.1$c.

The detectability of BH-NS mergers is much harder to predict. First
due to their virtually unconstrained rate and second since the
properties of the outflow are less certain. The latter can be
significantly improved by current and future merger simulations that
put focus on the ejected mass. In any case the potential of these
mergers to throw out a considerable amount of energy, $\sim 10^{51}$
erg, at mildly relativistic velocities can make them detectable to
their GW detection horizon, which is much further than the NS$^2$
horizon.

Finally, if compact binary mergers launch also collimated
ultra-relativistic outflows, and produce short GRB, then orphan
short GRB afterglows are also part of the post merger EM signal. The
detectability of radio orphan afterglows can be estimated based on
observations of short GRBs and is independent of whether they are
produced by mergers or not. We discuss their detectability in the
next subsection. All together, the range of the current predictions
is rather large, but with most parameters we expect detectable radio
signals. Some of the new generation radio telescopes have large
fields of view (e.g.,
ASKAP\footnote{http://www.atnf.csiro.au/projects/askap/technology.html}
with $30 {\rm ~deg}^2$ and Apertif with $8 {\rm ~deg}^2$;
\citealt{Apertif}) and improved sensitivities, making them ideal for
large scale sub-mJy blind survey. The EVLA, which has a smaller
field of view but a remarkable sensitivity, is the best facility for
GW triggered search. It is also currently the fastest radio-survey
instrument and it can carry-out a sub-mJy blind survey. All these
observatories have a very good chance to detect compact binary
merger remnants in dedicated blind searches. In fact with very
reasonable parameters (e.g., $E \sim 10^{50}$ erg of mildly
relativistic ejecta) a sub-mJy whole sky survey can detect thousands
of binary-merger radio remnants.

\subsection{Radio orphan afterglows of short GRBs}
Short GRB outflow begins highly relativistic and probably highly
beamed. Eventually it slows down (see \S \ref{sec Theory}) and
become detectable from all directions. Therefore, the rate estimate
equation  \ref{eq rate}
 is also  applicable for radio orphan afterglows when $\beta_0 = 1$. However
some of the parameters in equation \ref{eq rate} are not directly
observable. The observed quantities are isotropic equivalent
$\g$-ray energy, $E_{\g,iso}$, and the rate of bursts that point to
the observer ${\cal R}^{SHB}_{obs}$, while equation \ref{eq rate}
depends on $E = E_{iso} f_b$ and ${\cal R^{SHB}} = {\cal
R}_{obs}^{SHB} f_b^{-1}$, where $f_b<1$ is the fraction of the $4
\pi$ steradian covered by the jet and $E_{iso}$ is the isotropic
equivalent energy in the afterglow blast wave. X-ray observations
indicate that $\g$-ray emission in short GRBs is very efficient and
that in general $E_{iso} \sim E_{\g,iso}$ \citep{Nakar07}. We assume
that this is the case in the following discussion.

$E_{\g,iso}$ of short GRBs ranges at least over four orders of
magnitude ($10^{49}-10^{53}$ erg). The rate of observed short GRBs
is dominated by $10^{49}$ erg bursts, and the luminosity function
can be well approximated by a power-law, at least in the range $\sim
10^{49}-10^{51}$ erg, such that ${\cal R}_{obs}^{SHB}(E) \sim 10
E_{49}^{-\alpha} {\rm ~Gpc^{-3}~yr^{-1}}$ where $\alpha \approx
0.5-1$ \citep{NGF06,GP06}. Plugging these into equation \ref{eq
rate} we obtain
\begin{equation}\label{eq orphan_rate}
    N_{all-sky}^{SHB}(1.4 {\rm~ GHz}) \approx 1 f_b^{5/6} E_{49}^{\frac{11}{6}-\alpha} n^\frac{9p+1}{24} \epsB^\frac{3(p+1)}{8}
    \epse^\frac{3(p-1)}{2}  F_{lim,-1}^{-3/2} ~.
    \label{eq:orphan}
\end{equation}
This equation is similar to equation 9 of \cite{Levinson+02}, with
the observed luminosity function already folded in.

Narrower beamed bursts (with lower $f_b$) are more numerous and they
produce less total energy per burst.  The positive dependence of
equation \ref{eq orphan_rate} on $f_b$ implies that overall the
lower energy is ``winning" over the increased rate, and the
detectability of narrower bursts is lower. Using, equation \ref{eq
orphan_rate} we can put a robust upper-limit on the orphans rate
since all the parameters are rather well constraint by observations,
with the exception of $f_b$ which is $<1$ by definition. Therefore,
assuming that short GRBs are beamed, the detection of the common
$\sim 10^{49}$ erg bursts in a blind survey, even with next
generation radio facilities, is unlikely \citep{Nakar07}. However,
brighter events should be detectable. If the beaming is energy
independent, detectability increases with the burst energy.  The
luminosity function possibly breaks around $10^{51}$ erg, in which
case the orphans number is dominated by $10^{51}$ erg bursts. For
$f_b^{-1}=30$ we expect, from these bursts $\sim 10$ orphan
afterglows at a 0.1 mJy in a single $1.4$ GHz whole sky snapshot.

So far we discussed detectability in a blind survey. A followup
dedicated search would be, of course, more sensitive. If compact
binary mergers produce short GRBs than the energy of most GW
detected bursts will be faint with $E_{\g,iso} \sim 10^{49}$ erg.
The chance to detect their orphan afterglows again depended on their
total energy and thus on $f_b$. Equation \ref{eq Fpeak reg1} shows
that if $f_b^{-1}=30$ then detection should be difficult but
possible in a dedicated search mode. Note that since the energy of
the burst is low, the radio emission will evolve quickly, reaching a
peak and decaying on a week time scale, so a prompt and rather deep
search will be needed.

\subsection{Possible candidates of compact binary merger remnants}
\cite{Bower07} carried out a 5 GHz survey looking for transients on
timescales of a week to a year. The survey sensitivity for
transients with variability scale $<7$ day is 0.37 mJy with an
effective are of $\approx 10 {\rm ~deg^2}$. Events with variability
time scale of two months where surveyed at sensitivity of 0.2 mJy
with an effective area of $\approx 2 {\rm ~deg^2}$.

\cite{Bower07} report the detection of 10 transients. The most
interesting of those, in our context, is RT 19870422, which has a
variability time scale of two months. It is found within a star
forming galaxy at a distance of 1.05 Gpc, but at a significant
offset from the host nucleus. Its luminosity and time scale are
those expected from a $\sim 10^{50}$ erg  mildly relativistic
outflow that propagates in the ISM. It is, therefore, a prime
candidate for a compact binary merger radio remnant. Based on this
single event \cite{Bower07} infer a best estimate rate of $4 \times
10^3 {\rm ~Gpc^{-3}~yr^{-1}}$ for RT 19870422-like events. Taking a
$2\sigma$ Poisson error, the best estimate translates to a range of
$80 - 20,000 {\rm ~Gpc^{-3}~yr^{-1}}$, fully consistent with the
estimates of compact binary mergers. If this is indeed a merger
remnant then, since for optically thin spectrum the fluxes at 1.4
GHz and 5 GHz are not very different, a sub-mJy 1.4 GHz whole sky
survey would detect hundreds to thousands of radio remnants.
\cite{Bower07} suggested that this transient is a radio SNe similar
to SN 1998bw, but brighter. This is certainly a viable possibility,
however, if true then this radio SN is brighter by an order of
magnitude than the brightest radio SN ever observed before.
Unfortunately, lacking optical search for a SN or a multi-wavelength
measurement that determines the transient spectrum, it is impossible
to rule out any of the two possibilities.

An additional interesting candidate is RT 19840613. It is variable
on less than 7 days and it has a host galaxy at a distance of 140
Mpc. Even assuming a variability time scale of 7 days it is marginal
as a merger remnant candidate. Assuming 7 days variability
\cite{Bower07} find a rate of $(0.6-150) \times 10^4  {\rm ~Gpc^{-3}
~yr^{-1}}$, which is again only marginally consistent with current
estimates of compact binary merger rates. Therefore, while this may
be a merger remnant it is not a very promising candidate.
\cite{Bower07} suggest that this is also a SN 1998bw-like event.
While this possibility cannot be ruled out, the inferred rate is at
least an order of magnitude larger than that of 98bw-like events
(note that radio SNe as bright as SN 1998bw are very rare compared
to typical radio SNe). The additional 8 events detected by
\cite{Bower07} have no clear host galaxies and are therefore
probably not merger remnants.

\subsection{Identification and contamination}

A key issue with the detection of compact binary merger remnants in
blind surveys is their identification.  \cite{Ofek+11} present a
census of the transient radio sky. Luckily the transient radio sky
at $1.4$ GHz are relatively quite. The main contamination source are
radio active Galactic nuclei (AGNs), however their persistent
emission is typically detectable in other wavelength and/or deeper
radio observations. Moreover, the signal from a compact binary
merger is expected to be located within its host galaxy (otherwise
the density is too low), but away from its center. The host and the
burst location within it, should be easily detectable at the
relevant distances.

The only known, and guaranteed, transient 1.4 GHz source with
similar properties are radio SNe. Among these typical radio SNe are
the most abundant. Transient search over 1/17 of the sky with
$F_{lim}=6$ mJy at 1.4GHz \citep{Levinson+02,Gal-Yam+06,Ofek+10}
finds one radio SN. This rate translates to $10^{3}-10^{4}$ SNe in a
whole sky $F_{lim}=0.1$ mJy survey. These contaminators can be
filtered in three ways. First, by detection of the SN optical light.
However, the optical signal may be missed if it is heavily
extincted, and given the large number of radio SNe, misidentifying
even a small fraction of them may render the survey useless for our
purpose. The second filter is the optically thick spectrum at high
radio frequency ($\sim 10$ GHz) at early times, which is a result of
the blast wave propagation in a wind. Thus, a multi-wavelength radio
survey can identify radio SNe. The last filter is the
luminosity-time scale relation of typical radio SNe that is induced
by the outflow velocity \citep[e.g., figure 2 in][]{Chevalier+06b}.
Type II SN outflows are slow, $\sim 0.01$c, and therefore their
radio emission is longer/fainter than that expected for merger
remnants. The common type of Ib/c radio SNe is produced by $\sim
0.2$c blast waves but with much less energy than what we expect from
a binary merger outflow, and therefore their radio emission is much
fainter. The combination of any two of these filters will hopefully
be enough to identify all the typical radio SNe.

Slightly different contaminators are GRB associated SNe . Their
outflows is as fast and as energetic as those that we expect from a
binary merger and therefore their radio signature is similar in time
scales and luminosities. SN1998bw-like events are detectable by a
0.1 mJy survey at 1.4 GHz up to a distance of several hundred Mpc
for 40 days and their rate is $40-700 {\rm ~Gpc^{-3} ~yr^{-1}}$
\citep{skn+06}, implying at least several sources at any whole sky
snapshot. Here only the first filter (SN optical ligth) and possibly
the second (optically thick spectrum) can be applied. However, given
the high optical luminosity of GRB associated SNe and their
relatively low number this should be enough in order to filter them
out. These contaminators highlight the importance of a multi-wave
length strategy where an optical survey accompany the radio survey
to best utilize both surveys detections.

The results of \cite{Bower07} implies that thousands of sources with
properties similar to RT 19870422 are expected in a 5 GHz sub-mJy
all sky survey. If these events are optically thick during their
whole evolution than these are not merger remnants and they could be
easily filtered out. Moreover, in that case their rate in a 1.4 GHz
survey should be lower by two orders of magnitude. If these events
show a synchrotron optically thin spectrum and no optical
counterparts, then they should be abundant also in a 1.4 GHz survey,
and their origin is most likely compact binary mergers.

Finally, radio is the place to look for blast waves in tenuous
mediums, regardless of their origin. Any source of such explosion,
being a binary merger, a GRB or a SN, produces a radio signature.
Therefore, all the strong explosions may be detectable is a deep
radio survey, this include for example long GRB on-axis and off-axis
afterglows and giant flares from extra galactic soft
gamma-repeaters. The difference between the radio signatures of the
different sources (amplitude, spectrum and time evolution) depends
on the blast wave energy and velocity and on the external medium
properties. We thus will be able to identify the characteristics of
binary mergers outflows. If, however, there is a different source of
$\sim 10^{50}$ erg of mildly relativistic outflow that explodes in
the ISM it will be indistinguishable from binary mergers (at least
not in the radio). Currently we are not aware of any such source,
with the exception of long GRBs at the low end of the luminosity
function, but these are too rare to contaminate a survey. Any other
source of such outflows, if exist, will probably be a part of the
family of collapsing/coalesing compact objects.


\section{Conclusion}

Compact binaries are expected to eject sub-relativistic, mildly
relativistic and possibly ultra-relativistic outflows as part of
their merger process. We have shown that these outflows will
inevitably produce a long lived radio remnant. These are the most
robust predictions of an EM counterpart of the merger GW signal. The
radio remnant appears weeks to years after the merger and remains
bright for a similar time. Therefore, a trigger following a
detection of GW signal can wait for a week after the event and no
online triggering is needed. In addition the long lasting remnants
enable a detection in a blind survey. For mildly relativistic
outflows with $10^{49}$ erg that propagate in the ISM we expect a
few weeks radio transients with a 1.4 GHz flux of  $\sim$ 0.3 mJy
from sources at $\sim 300$ Mpc, the advanced LIGO-Virgo horizon for
NS$^2$ mergers.
The BH-NS GW horizon is farther, but current numerical simulations
suggest they involve higher energy outflow  resulting in a
comparable flux. Follow up observations of GW candidate events, at a
level of $\sim 10 \mu$Jy are feasible and are very likely to show a
radio transient for either NS$^2$ merger of BH-NS merger.

We find that the optimal frequency to carry out a search for merger
remnants is $1.4$ GHz. Assuming a mildly relativistic outflow with
$10^{49}$ erg the canonical NS$^2$ merger rate of 300 Gpc$^{-3}$
yr$^{-1}$(and a range of 20 - $2 \times 10^4$ Gpc$^{-3}$ yr$^{-1}$)
implies a detection of $\sim 20$ (1-1200 correspondingly) radio
NS$^2$ remnants in a $0.1$ mJy all sky survey. This rate depends
quadratically on the outflow energy, so a very plausible ejected
energy of $10^{50}$ erg increases the rate by two orders of
magnitude, making them detectable even in a survey that covers only
a small fraction of the sky or that is at a mJy sensitivity.
Therefore carrying out a large field-of-view and sensitive GHz
survey by currently available facilities has a great potential to
constrain the rate of binary mergers, a piece of information  that
is of great importance for the design and operation of GW detectors.

Even if mergers do not launch a significant mildly relativistic
ejecta they are still expected to produce an energetic ($\gtrsim
10^{50}$ erg) sub-relativistic (0.1-0.2)c outflows. These outflows
will also produce radio remnants. These remnants will be fainter,
detectable only to a distance of $\sim 100$ Mpc at 0.1 mJy, and will
evolve more slowly, on time scales of 3-10 yr. These transients are
also detectable at a rate of $\sim 10$ over the whole sky at 0.1
mJy, although identifying them as transients is harder and it
requires a long term survey.

We estimate the detectability of short GRB orphan afterglows, which
may also be produced by compact binary mergers if they are launching
also ultra-relativistic outflows. These estimates are based on short
GRB observations and are therefore indifferent to whether short GRBs
are binary mergers or not. The main uncertainty in the rate
estimates is the GRB beaming factor. We find that assuming
$f_b^{-1}=30$ there are expected to be about 10 orphan afterglows at
a 0.1 mJy in a single $1.4$ GHz whole sky snapshot. The duration of
these afterglows is several weeks. If binary mergers are short GRBs
than a GW triggered event will most likely be of a low energy GRB,
$E_{\g,iso} \sim 10^{49}$ erg, and a true energy, after beaming
correction, that is even lower. The radio orphan afterglow will
probably still be detectable in a deep search. However its
variability time scale is short, about a week, so the search should
be done promptly.

Remarkably, the observed 5 GHz transient RT 19870422, detected by
\cite{Bower07} fits very well our estimates of the typical expected
properties of a compact binary merger radio remnant.   At a distance
of 1 Gpc and a duration of two months this transient is what
expected from a mildly relativistic outflow with $\sim 10^{50}$ erg.
The rate inferred from this single event is also fully consistent
with that of NS$^2$ mergers. This transient is an excellent
candidate to be the first observed radio remnant of a merger.
Unfortunately, one can not rule out the possibility,  suggested by
\cite{Bower07}, that this is an especially bright radio SN. Note,
however, that this  interpretation requires a SN  brighter  by an
order of magnitude than any radio SNe previously observed.
Simultaneous optical observations or multi-wavelength radio
observations could have  easily distinguished between the two
possibilities. The first could have determined if there was a SN or
not. The second could have distinguished between an optically thick
radio spectrum expected in radio SNe vs the optically thin spectrum
expected in merger remnants at these frequencies at all times.
Unfortunately, no such observations were available. However, the
rate implied by this even is very high and similar events should be
detected in a sub-mJy survey of even a small fraction of the sky.
Therefore, the nature of this type of events can be easily probed
with current facilities.



Our results show the great potential of  $1.4$ GHz radio transient
observations at the sub-mJy level for the detection of NS$^2$
mergers. On the observational side these predictions provide an
excellent motivation for carrying out a whole sky sub-mJy survey
using the EVLA or  other upcoming radio telescopes. The main source
of contamination in such surveys would be radio supernova and those
could be distinguished from compact binary mergers by their optical
signal, spectrum and other characteristic properties.

While it is clear that compact binary mergers produce
sub-relativistic to relativistic outflows, details of those outflows
are not well determined at present. This is to large extent because
of lack of interest rather than because of specific difficulties in
analyzing their properties. Our analysis elucidate  the importance
of a detailed quantitative estimates concerning these outflows, a
task that is within the scope of current simulations.

We thank Dale Frail, Shri Kulkarni, Andrew MacFadyen, Eran Ofek and
Stephan Rosswog for helpful discussions. This research was supported
by an ERC advanced research grant, by the Israeli center for
Excellent for High Energy AstroPhysics, by the Israel Science
Foundation (grant No. 174/08) and by an IRG grant.


\end{document}